\documentstyle[prb,aps]{revtex}
\begin{document}
\draft
\title{Analytical solution for the Fermi-sea
energy of two-dimensional electrons in a magnetic field:
lattice path-integral approach and quantum interference}
\author{Franco Nori and Yeong-Lieh Lin}
\address
{Department of Physics, The University of Michigan, Ann Arbor, MI 48109-1120}
\maketitle

\begin {abstract}
We derive an exact solution for the total kinetic energy of
noninteracting spinless electrons at half-filling in two-dimensional
bipartite lattices. We employ a
conceptually novel approach that maps this problem exactly into a
Feynman-Vdovichenko lattice walker. The problem is then reduced to the
analytic study of the sum of magnetic phase factors on closed paths.
We compare our results with the ones obtained through numerical
calculations.

\end{abstract}

\twocolumn

\narrowtext

{\it Introduction.---} Non-interacting tight-binding electron models
at half-filling in
two-dimensional (2D) bipartite (e.g., square and hexagonal)
lattices have recently received renewed attention due to their role in
condensed matter\cite{haldane} and particle physics\cite{semenoff}.   For
instance, several quantum field theories\cite{semenoff}
arise in a natural way from 2D tight-binding lattice fermion problems
{\it at half filling\/} for hexagonal\cite{haldane,semenoff}
(e.g., 2D graphite\cite{graphite}) and square
 lattices\cite{semenoff,fradkin1}.
These quantum field theories\cite{semenoff} are important to the problem
of dynamical symmetry breaking, which plays a central role in
many current areas of research; for instance, they provide a
possible mechanism for generating the fermion mass spectrum
in elementary particle physics\cite{jackiw}.  Furthermore,
noninteracting 2D tight-binding electrons in a perpendicular
magnetic field have been the subject of intense study
in areas of current interest like mesoscopic structures and the
quantum Hall effect.
More recently, the behavior of the kinetic energy of a 2D
non-interacting electron gas under the influence of both a periodic
potential and a magnetic field
has been analyzed by Hasegawa et al.\cite{hlrw} and
many others\cite{hasegawa2,lpr}.  These results
have been related to mean-field approaches to the $t-J$ model\cite{lpr}
of high-temperature superconductors.
It is of value to obtain analytical results for this important problem
that has recently motivated many perturbative and numerical studies.

The goal of this paper is to present exact results that relate the kinetic
energy of the half-filled Fermi sea of tight-binding electrons with
sum-over-paths on the lattice. The problem is then reduced to the study of
phase factors on closed paths. From the computational point of view, we would
like to present an alternative to the standard approaches. From a physical
point of view, and following Feynman's program, we would like to
reformulate this quantum problem as an ``average over histories''.

Recently, the lattice path approach has also been applied
to two very diverse problems; the computation of:
(1) equilibrium crystal shapes\cite{holzer}, and
(2) the superconducting transition temperature in wire
micronetworks and Josephson-junction arrays\cite{nn}.

{\it Fermi-sea energy.---}   The kinetic energy of spinless electrons on a
2D lattice
in a uniform magnetic field is described by the Hamiltonian $H=- \sum_{<ij>}
c_{i}^{+}  c_{j}\   \exp [i \phi_{ij}]$, where $<ij>$ refers to
nearest-neighbor sites and
$ \phi_{ij}=2\pi\int_{i}^{j} {\bf A\cdot}d{\bf l}$
in units of the flux quantum.  Also, $|j>\equiv c_{j}^{+}|0>$
defines a localized state centered at site $j$, and we will
work on the $\{|j>\}$ basis.  The above
Hamiltonian has a fractal quantum energy
spectrum as a function of an applied magnetic field and also plays a central
role in the physics of
particles with fractional statistics (anyons)\cite{girvin}.

At half-filling, the Fermi-sea ground-state energy is the sum of
the lowest $N/2$ eigenvalues, where $N$ is the total number of
sites. Since the energy spectra of bipartite (e.g., square and
hexagonal) lattices are symmetric under $\{E\} \rightarrow \{-E\}$,
we can write the Fermi-sea ground-state energy per site
as $E_{T}=\frac{1}{N}\sum_{E<0} E=-(z/2N)\ {\rm Tr}\ |H_{0}/z|$. Here, $H_{0}$
is the corresponding diagonalized Hamiltonian obtained from $H$ by a similarity
transformation. Also $z$ is the coordination number of the lattice and Tr
denotes the trace. For a square (honeycomb) lattice, $z=4$ ($3$). Note that the
energy spectrum of a lattice is bounded between $-z$ and $z$. By expanding
$|x|\equiv|H_{0}/z|$ into a Chebyshev series $T_{2k}(x)$, it follows that
$$|x|=\frac{2}{\pi}+\frac{4}{\pi}\sum_{k\geq1}\frac{(-1)^{k+1}}{4k^{2}-1}T_{2k}(x).$$
Also, using the identity
$$T_{2k}(x)=(-1)^{k}k\sum_{l=0}^{k}(-1)^{l}\frac{(k+l-1)!}{(k-l)!(2l)!}(2x)^{2l},$$
we obtain
\begin{equation}
E_{T}=-\frac{z}{N\pi}\{1-2\sum_{k\geq1}\frac{k}{4k^{2}-1}[\sum_{l=0}^{k}\Gamma_{l} {\rm Tr}(H^{2l})]\} ,
\end{equation}
where $$\Gamma_{l}=(\frac{-4}{z^{2}})^{l}\frac{(k+l-1)!}{(2l)!(k-l)!}.$$
We have replaced ${\rm Tr}(H_{0}^{2l})$ by
${\rm Tr}(H^{2l})$, as they are equal to each other.

Now let us examine the term ${\rm Tr}(H^{2l})$ more closely. Assuming
periodic boundary conditions on the lattices, we have
\begin{eqnarray}
\lefteqn{{\rm Tr}(H^{2l})}  \nonumber \\
& &=<1|H^{2l}|1>+<2|H^{2l}|2>+ \cdots +<N|H^{2l}|N> \nonumber \\
& &= N<j|H^{2l}|j>.
\end{eqnarray}
 The Fermi-sea ground-state energy is then exactly given by
\begin{equation}
E_{T}(\Phi)=-\frac{z}{\pi}\{1-2\sum_{k\geq1}\frac{k}{4k^{2}-1}[\sum_{l=0}^{k}\Gamma_{l}M_{2l}(\Phi)]\},
\end{equation}
where $\Phi$ equals $2\pi$ times the flux through each lattice plaquette, and
$M_{2l}(\Phi)\equiv <j|H^{2l}|j>$ is a moment, or lattice path-integral,
discussed in more detail below.

We have also studied two additional expansion schemes; one of them using a
different set of orthogonal polynomials and the other based on a power series.
One of them exploits the expansion of $|x|$ in terms of Legendre polynomials,
$P_{2k}(x)$, as
$$|x|=\frac{1}{2}+\sum_{k\geq1}\frac{(2k-3)!!}{(2k+2)!!}(4k+1)P_{2k}(x).$$
After expressing each $P_{2k}(x)$ in terms of a power series, it is clear that
\begin{eqnarray}
\lefteqn{E_{T}(\Phi)} \nonumber \\
& &
=-\frac{z}{4}\{1-2\sum_{k\geq1}\frac{4k+1}{4^{k}}\frac{(2k-3)!!}{(2k+2)!!}[\sum_{l=0}^{k}\Omega_{l}M_{2l}(\Phi)]\},
\end{eqnarray}
where $$\Omega_{l}=(\frac{-1}{z^{2}})^{l}\frac{(2k+2l)!}{(2l)!(k-l)!(k+l)!},$$
The other scheme starts by rewriting  $E_{T}$ as follows
$$E_{T}=-\frac{1}{2N}{\rm Tr}[(H_{0}^{2})^{1/2}]=
-\frac{a}{2N}{\rm Tr}\{[1+(-1+\frac{H_{0}^{2}}{a^{2}})]^{1/2}\},$$
where $a$ is a constant to ensure $|-1+H_{0}^{2}/a^{2}|<1$.
In general, $a=2\sqrt{2}$ for a square lattice, and $a=3/\sqrt{2}$ for a
honeycomb lattice. However, for a magnetic flux near $1/2$, we find that
$a=2$ produces better results for both lattice types, since the
energy density vanishes for larger values of $E=2\sqrt{2}$. By direct series
expansions, we then obtain
\begin{equation}
E_{T}(\Phi)=\frac{a}{2}\{-1+\sum_{k\geq1} \Lambda_{k}[\sum_{l=0}^{k}
\frac{(-1)^{l}k!}{(k-l)!l!}\frac{1}{a^{2l}} M_{2l}(\Phi)]\}.
\end{equation}
with $$\Lambda_{k}=\frac{1}{2^{2k-1}k} \frac{(2k-2)!}{[(k-1)!]^{2}}.$$

Among the above three solutions for $E_{T}(\Phi)$ (namely, Eqs.~(3-5)), Eq.~(3)
gives the best results because of its relatively rapid convergence. We will
return to the discussion on their accuracy later.

{\it Sum over paths.---} The lattice path-integral (or moment) used in this
work is
\begin{equation}
M_{2l}  \; \equiv \; <j|H^{2l}|j> \ =
\sum_{All \ 2D \ closed \ paths} \
e^{i\Phi_c} \ \ ,
\end{equation}
\noindent
where $\Phi_c/2\pi$ is equal to the net flux enclosed by the
directed closed path and $2l$ is referred as its order. The physical meaning of
the above
quantum mechanical expectation value is simple.  The
Hamiltonian {\bf H\/} is applied $2l$ times
to the initial state $|j>$ localized at site $j$.  This provides
enough kinetic energy for the electron to hop through $2l$ bonds,
reaching the new state $H^{2l}|j>$ located at the end of the
path.  The above expectation value
is non-zero only when the path ends at the starting site $j$.
In our problem, quantum interference arises because the phase factors
of different closed paths, or separate contributions from the same
path, interfere with each other, sometimes producing cancelations
in the phases.   Therefore, our calculations generalize the
Aharonov-Bohm phase factor, obtained by having an electron going around a
single
flux-enclosing-loop, to the (multiply-connected) lattice case.

The basic problem is now reduced to the computation of the lattice
path-integrals.  This is a very difficult task, since each moment involves an
enormous number of different loops, each one weighted by its corresponding
phase factor.  We have considerably simplified this calculation by analyzing
the symmetries of the problem.  In the next few paragraphs, we will list
the most important symmetries involved and we will present a few
examples of how the method works.  Further details, with applications
of this method to other problems, will be presented elsewhere.

We will now compute $M_{2l}$. We consider a unit spacing for square as well as
graphite lattices and employ the Landau gauge ${\bf A}=(0,Bx)$.
Note that $M_{0}$ equals $1$ and is independent of the type of lattice. Also,
the $M_{2l}$'s are gauge invariant. First let us investigate the square
lattice.  Writing the coordinates
of site $j$ as $j=(m,n)$, we define an auxiliary quantity, $W_{r}(m,n)$,
which is the sum over all possible paths of $r$ steps on which an electron
may hop from some given site to the site $(m,n)$. From the definition of
$W_{r}(m,n)$, it is evident that it obeys the recurrence relation
\begin{equation}
W_{r+1}(m,n)=W_{r}(m\pm 1,n)+ e^{\mp im \Phi}W_{r}(m,n\pm 1).
\end{equation}
Equation~(8) states that the site $(m,n)$ can be reached by taking the
$(r+1)$th step from the four nearest-neighbor sites. The factors in front
of the $W_{r}$'s account for the presence of the magnetic field. We can
construct further recurrence relations successively. For example,
\begin{eqnarray}
\lefteqn{W_{r+2}(m,n)=4W_{r}(m,n)+(1+e^{\pm i\Phi})}  \nonumber \\
& &\times[e^{im \Phi}W_{r}(m\pm 1,n-1)+e^{-im \Phi}W_{r}(m\mp 1,n+1)]
\nonumber \\    & &+ W_{r}(m\pm 2,n)+ e^{\pm 2im \Phi}W_{r}(m,n\mp 2).
\end{eqnarray}

Examining the action of the Hamiltonian on the state $|j>=|m,n>$, we find that
\begin{equation}
-H |m,n>=|m\pm 1,n>+e^{\mp im \Phi}|m,n\pm 1>.
\end{equation}
Hence, by comparing Eq.~(7) with Eq.~(9), we obtain $M_{2l}$ which is just
the coefficient of $W_{r}(m,n)$ in the recurrence relation for
$W_{r+2l}(m,n)$. This coefficient is obviously
{\em the sum over all possible paths} which return an electron to its
original site $(m,n)$ after hopping $2l$ steps.  Each path has a
phase factor corresponding to the net flux going through the
{\it directed\/}
(e.g., $-\Phi_c$ clockwise and $\Phi_c$ counter-clockwise)
path.

It is worthwhile to notice the following symmetries when constructing
recurrence relations and obtaining  $M_{2l}$.

\ (a) The recurrence relation for $W_{r+l}$ contains only terms
$W_{r}(m+p,n\pm q)$ and $W_{r}(m-p,n\pm q)$ which satisfy the restriction :
$p+q=1,3, \ldots ,l$, for $l$ odd and = $0,2, \ldots ,l$, for $l$ even.

\ (b) The coefficients in front of the $W_{r}$'s can be factored into two
parts. Each multiplicative factor involving coordinate $m$ is always
of the form $e^{\mp iqm \Phi}$, for $W_{r}(m+p,n\pm q)$ and
$W_{r}(m-p,n\pm q)$. We shall refer to the rest of the prefactors (the part not
involving $m$) as $C_{r}$. For instance, $C_{r}(m\pm 1,n-1)=C_{r}(m\mp
1,n+1)=1+e^{\pm i\Phi}$ in the above expression (Eq.~(8)) for $W_{r+2}(m,n)$.
These $C_{r}$'s satisfy
$ C_{r}(m+p,n-q)=C_{r}(m-p,n+q)=C_{r}(m+q,n-p)=C_{r}(m-q,n+p)$ and
$C_{r}(m+p,n+q)=C_{r}(m-p,n-q)=C_{r}(m+q,n+p)=C_{r}(m-q,n-p).$
       It can be shown that the latter set becomes equivalent to the former
one when $\Phi \rightarrow -\Phi$ (and vice-versa).

\ (c) To obtain $M_{2l}$, it is sufficient to compute the complete
recurrence relation for $W_{r+l}(m,n)$.

We have computed the path-integrals up to $M_{40}$. Here we list $M_{2}$
through $M_{10}$: $4$, $28+ 8\cos\Phi$, $232+ 144\cos\Phi+ 24\cos2\Phi$, $2156+
2016\cos\Phi+ 616\cos2\Phi+ 96\cos3\Phi+ 16\cos4\Phi$, $21944+ 26320\cos\Phi+
11080\cos2\Phi+ 3120\cos3\Phi+ 840\cos4\Phi+ 160\cos5\Phi+ 40\cos6\Phi$, for
$l=2,\cdots,10$. Notice that moments with odd orders are always zero because
there is no path with an odd number of hops for which the electron may return
to the initial site on a bipartite lattice.

We now consider the hexagonal lattice, which consists of two
interpenetrating triangular sublattices.  Following similar techniques,
we have two different formulae for $W_{r+1}(m,n)$. The proper choice between
them depends on the sublattice to which the site $(m,n)$ belongs. However, both
choices lead to the same results for the path-integrals.  These have been
computed up to $M_{60}$. We list $M_{2}$
through $M_{10}$ here: $3$, $15$, $87+6\cos\Phi$, $543+ 96\cos\Phi$, $3543+
1080\cos\Phi+ 30\cos2\Phi$, for $l=2,\cdots,10$.

In general, the path-integrals can be analytically computed by hand to any
desired order through the techniques discussed above. We have computed by hand
the moments up to the $20$th order for both square and hexagonal lattices.
However, these and the higher order moments can be most conveniently obtained
by using computer symbolic-manipulation software. The correctness of the
calculated moments is assured by the consistency of the results obtained by
hand and by computer.

{\em Fermi-sea energy values and discussion.---}After computing the
lattice path-integrals, we can now proceed to calculate the kinetic
energy of the half-filled Fermi sea. Recall that we have obtained the moments
up to $M_{40}$ ($M_{60}$) for the square (honeycomb) lattice. Therefore,
by truncating the series at $k=20$ ($30$) for a square (honeycomb)
lattice in Eqs.~(3-5), we obtain analytic closed-form expressions for
the ground state energy as an explicit function of the magnetic flux.
In Table I, we present our results for the Fermi sea energies at
various values of the flux by using Eqs.~(3,4). For the square (hexagonal)
lattice, the results are
obtained by using the path-integrals up to $M_{20}$, $M_{30}$ and $M_{40}$
($M_{20}$, $M_{40}$ and $M_{60}$) respectively. Here, instead of showing the
exact numerical expressions (involving, {\em e.g.}, $\pi$'s and square roots of
integers), we present the actual numerical values for easier comparison
purposes. It is worthwhile to notice that results obtained by using moments up
to $M_{20}$, are already in excellent agreement with those obtained in
Refs.~6,7. The values obtained by using higher order moments are essentially
identical. Also Eq.~(3) and Eq.~(4) produce almost identical results. Although
the results obtained from Eq.~(5) are not as close as those obtained from
Eq.~(3,4), they are consistent within $\pm 0.03$.

It should be pointed out here that
the tight-binding model $H$ does not include the diamagnetic energy of the
tight-binding orbitals
and the reduction of the hopping amplitude by the magnetic field.  This
issue is outside the scope of this paper.  The interested reader can find
a very detailed analysis of these points in Ref.~12 and references therein.
We also note that
an expression for the density of states, in terms of elliptic integrals
and obtained through a completely
different approach, has been known for some
time and used, for instance, in Ref.~7. Also, several
groups\cite{hlrw,hasegawa2}, including ours, have obtained results by other
methods, including purely numerical approaches. Furthermore, different
approaches on similar problems have been recently explored\cite{pernici}.

In conclusion, the theory of electronic diamagnetism in
two-dimensional lattices has been studied extensively due to its many
applications in very diverse areas of physics.  In particular,
several computations of the Fermi-sea energy have recently
attracted considerable attention by many workers.
We use a conceptually novel approach that maps the problem exactly
onto a Feynman-Vdovichenko lattice walker.  More specifically, we
derive an expression for the Fermi-sea kinetic energy at half-filling,
as a function of a uniform perpendicular magnetic field, in terms of
the quantum interference originating from the sum over 2D lattice closed paths,
each loop weighted by the phase
factor corresponding to the net flux enclosed. The energies obtained are
essentially identical to the ones obtained through numerical calculations.
We have shown that
lattice path-integral techniques can be successfully applied to this
system and we expect this approach to be applicable
to many other electronic problems.

FN acknowledges conversations with M. Pernici. This work has been supported in
part by the NSF grant DMR-90-01502.

\widetext
\begin{table}
\caption{Fermi-sea ground-state energies at half-filling for the square and
hexagonal lattices and for various values of the flux. The first three (4th to
6th) rows  present results obtained from Eq.~(3) (Eq.~(4)), {\em i.e.}, by
using a Chebyshev (Legendre) series expansion. The numbers in parentheses
indicate
the highest order of the path-integral used. They are compared with the
numerical results (Num.) from Refs.~6,7.}
\begin{tabular}{rccccccc}
\multicolumn{8}{c}{Square lattice}  \\
$\frac{\Phi}{2\pi}$   &$0$  &$1/8$  &$1/6$  &$1/4$ &$1/3$  &$3/8$  &$1/2$   \\
 \hline
C~(20)  &$-0.8114920$  &$-0.8254256 $ &$-0.8288598 $ &$-0.8592316 $
&$-0.8569940 $ &$-0.8801229$ &$-0.9583405$  \\
C~(30)  &$-0.8109902$  &$-0.8231301 $ &$-0.8347727 $ &$-0.8599869 $
&$-0.8576158 $ &$-0.8775567$ &$-0.9581710$  \\
C~(40)  &$-0.8108108$  &$-0.8252121 $ &$-0.8361273 $ &$-0.8588387 $
&$-0.8574827 $ &$-0.8774209$ &$-0.9580550$  \\
L~(20)  &$-0.8114462$  &$-0.8250180 $ &$-0.8286112 $ &$-0.8603026 $
&$-0.8567193 $ &$-0.8799075$ &$-0.9584124$  \\
L~(30)  &$-0.8109779$  &$-0.8234147 $ &$-0.8348405 $ &$-0.8600546 $
&$-0.8576395 $ &$-0.8777049$ &$-0.9582486$  \\
L~(40)  &$-0.8108047$  &$-0.8254116 $ &$-0.8362933 $ &$-0.8587258 $
&$-0.8575003 $ &$-0.8774703$ &$-0.9580659$  \\
Num. &$-0.811$  &$-0.826 $ &$-0.835 $ &$-0.859 $ &$-0.857 $ &$-0.880$ &$-0.958$
\\ \hline \hline
\multicolumn{8}{c}{Hexagonal lattice} \\
$\frac{\Phi}{2\pi}$   &  $0$ &  &  $1/4$  &  &$1/3$ & &  $1/2$   \\   \hline
C~(20)  & $-0.7869927$ & &$-0.7506642 $ & &$-0.7489389$ & & $-0.7527022 $ \\
C~(40)  & $-0.7872330$ & &$-0.7527290 $ & &$-0.7506511$ & & $-0.7538837 $ \\
C~(60)  & $-0.7872775$ & &$-0.7530423 $ & &$-0.7509664$ & & $-0.7537718 $ \\
L~(20)  & $-0.7869492$ & &$-0.7512373 $ & &$-0.7486450$ & & $-0.7524125 $ \\
L~(40)  & $-0.7872308$ & &$-0.7527266 $ & &$-0.7506855$ & & $-0.7537969 $ \\
L~(60)  & $-0.7872783$ & &$-0.7530359 $ & &$-0.7510000$ & & $-0.7537457 $ \\
Num. & $-0.787$ & &$-0.753 $ & &$-0.751$ & & $-0.754 $ \\
\end{tabular}
\label{table2}
\end{table}


\begin{references}
\bibitem{graphite}
The 2D hexagonal (graphite or honeycomb) lattice describes the atomic
arrangement of carbon
atoms in graphite, a highly anisotropic semi-metal with a stacking structure
of almost independent hexagonal layers, each one with one valence electron
per site.  Given the double spin degeneracy this means a half-filled
band.  The electron dynamics is then modeled by a tight-binding model.
In this work, we will concentrate on the half-filled band case;
thus, the ground state has positive energy states empty,
and negative energy states filled.
\end{references}
\end{document}